# Tax Bond Creation Using a Structural Model and its Extensions

Suren Harutyunyan

*Pompeu Fabra University*

*suren.harutyunyn01@estudiant.upf.edu*

*Abstract*

*This article describes and explores taxes and debt in finance. Here a situation is thought about, where tax payments would qualify to be considered as debt. Using this principle we can infer that it is possible to create and price a type of bond (Tax Normalization Guarantee) for companies, which would allow them to enter in temporary tax breaks to allow them to free capital. Finally it is explored a way to structure these bonds in financial products and valuate them.*

## I. Introduction

In this work the author shows how theoretically companies with a large tax burden could use financial products in order to free capital and enter in temporary tax breaks. In order to achieve this it could issue a new product called TNG (Tax Normalization Guarantee), which would behave like a bond. The bond simply would help the companies to free capital and to enter in temporary tax breaks, in order to obtain more available cash for certain projects. Here, the main extensions of the Black-Sholes formula are used to be able to value these bonds.

Afterwards, the article considers a hypothetical situation where these TNG's could be structured, by third parties, in different financial products.
These derivatives could be priced using the model presented by Hull, Predescu, and White (2005). This would make possible to the broader public to gain an exposure to the cash-flows of these types of bonds and securitized products.

II. Theoretical example

Nowadays taxes (especially corporate income taxes) are an important issue for governments and for corporations alike. In this work it would be useful to think about taxes like a variable amount $T_1$ that a company $C_1$ has to disburse after a certain amount of time (t) to another institution $C_2$. That said we can think of taxes as some kind of liability, i.e. they can be treated as debt.

That said, we have to think in a situation where $C_1$ for some specific reasons needs to free capital. The problem is that $C_1$ cannot free $T_1$, because it consists of the total tax payouts from taxes due at time t.

To solve this we have to introduce a bond contract that here we will call TNG (Tax Normalization Guarantee). This would consist of a contract between two parties ($C_1$ and $C_3$), which $C_1$ will have the option to initiate in order to enter in a temporary tax break. The contract will have to establish that when $C_1$ initiates the TNG for some specific reasons $C_3$ will make the specified tax payouts on behalf of $C_1$, amount which will have to be returned by $C_1$ in a specified time in the future. For now this oversimplification will have to apply.

This way, the TNG's will become a type of bond, which we already know how to price. Also, if in turn $C_3$ creates a certain quantity of these issuing it could create a CDO (using the method of Section IV), whose underlying assets could consists of these bonds. After that, with techniques used in Section II it could price the default correlation of the tranches and sell them to $C_4$, using the proceeds to compensate for the amount that $C_2$ used to meet its obligations.

For the framework, we also have to remember to apply the assumptions made in Section III, especially A.2 and A.3.

### III. On the corporate debt pricing

*Risky discount bon pricing (3.1)*

Black and Scholes (1973] introduced to us a new equilibrium theory for valuing options, the attractiveness of which consisted in that the formula presented was based in variables that could be observed. The intuition told that this type of analysis could possible applied to other securities. Merton (1974) used this method to create the structural model of default for a single company. The basic assumptions made in the literature, see Merton (1974) and Ingersoll (1976), are as follows

A.1) Perfect market: the capital markets are perfect, so there are no transaction costs, no tax costs, no agency costs. Assets are perfectly divisible.

A.2) Every individual thinks that he can buy and sell as much of any security as he wants, and this cannot affect the market price

A.3) We have a market for borrowing and lending at a constant interest rate.

A.4) Individuals are allowed to have short positions on all assets, including the riskless asset. Full use of the proceeds of the short sales of all assets is allowed.

A.5) Trading in assets takes places continuously

A.6) The value of the firm does not vary to its capital structure, i.e. the Modigliani-Miller theorem holds.

A.7) The term structure is "flat" and non-stochastic. So the price of the riskless discount bond promising a payment of a dollar at a time $\tau$ in the future is $P(\tau) = \exp[-r\tau]$, being r the riskless rate of interest

A.8) The value of the firm, V, follow a diffusion-type stochastic process with differential equation

$$dV = (\alpha V - C)dt + \sigma V dz$$

where

$\alpha$ is the instantaneous expected rate of return on the firm per unit time. $\sigma^2$ is the instantaneous variance of the return on the firm per unit time. C, if positive, it is the total cash outflow paid out by the firm per unit time. If it is negative it is the net dollar received by the firm from new financing. dz is a Gauss-Weiner process. Under these same assumptions Merton (1974), also Black and Cox (1976), created a valuation equation. He showed that any contingent claim, F (V, t), whose value can be expressed as a function of asset value and time, must satisfy this general form

$$(1) \quad \frac{1}{2}\sigma^2 V^2 f_{vv} + \big(rV - p(V,t)\big)f_v - rf + f_t + p'(V,t) = 0$$

Here f is a name for the securities of the firm. V denotes the value, t denotes the time, $\sigma^2$ expresses the instantaneous variance of the return on the firm. $p(V,t)$ is the net total outflow made or it is the inflow received by V, and $p'(V,t)$ is the payout received or the payment made by the security f.

Also, if the claim has a maturity date $\tau$ periods in the future we can express it like F (V, $\tau$), and (1) would be

$$(2) \quad \frac{1}{2}\sigma^2 V^2 f_{vv} + \big(rV - p(V,\tau)\big)f_v - rf + f_\tau + p'(V,\tau) = 0$$

In his seminal paper Merton (1974) introduces a method for pricing corporate debt presenting us a formula that is based on inputs that are observable; he obtains the basic equation for the pricing of financial instruments from the valuation model proposed by Black and Scholes (1976), where $\sigma^2$ is constant. He establishes the value of debt issue as

$$(3) \quad F[V, \tau] = Be^{-r\tau} \left\{ \Phi[h_2(d, \sigma^2\tau)] + \frac{1}{d} \Phi[h_1(d, \sigma^2\tau)] \right\}$$

where

$$d \equiv Be^{-r\tau}/V$$

$$h_1(d, \sigma^2\tau) \equiv -\left[\frac{1}{2}\sigma^2\tau - \log(d)\right]/\sigma\sqrt{\tau}$$

$$h_2(d, \sigma^2\tau) \equiv -\left[\frac{1}{2}\sigma^2\tau - \log(d)\right]/\sigma\sqrt{\tau}$$

*Risky coupon bond pricing (3.2)*

But this is the simplest form of corporate bond (i.e. the discount bond where there are no coupon payments). For the perpetual risk coupon bond Merton (1974) used (3.1.1) and only modified the indenture condition to require continuous payments at a coupon rate per unit time $\bar{C}$. So using the identity F = V − f, the solution for the perpetual risk coupon bond was obtained as

$$(4) \quad F[V, \infty] = \frac{\bar{C}}{r} \left\{ 1 - \frac{\left(\frac{2\bar{C}}{\sigma^2 v}\right)^{\frac{2r}{\sigma^2}}}{\Gamma\left(2+\frac{2r}{\sigma^2}\right)} M\left(\frac{2r}{\sigma^2}, 2+\frac{2r}{\sigma^2}, \frac{2\bar{C}}{\sigma^2 v}\right) \right\}$$

The model explained that a default could occur if the value of the assets of the company is below the face value of the debt at particular future time.

This model afterward was accompanied by an important extension created by Black and Cox (1976). This work extended the model to see the effects of safety covenants on the value and the behavior of the firm's securities. Here was included afirst passage time structure, where a natural form of safety covenant is similar to this: the bondholders have the right to enforce the bankruptcy of the firm and obtain the ownership of the assets, if the value of the assets of the firm falls below a specified level. Rather than repeat the exact parts of the analysis we will focus on the broader framework that gives us the tools necessary to build the security.

## IV. On the credit derivative pricing

*CDO conceptual creation (4.1)*

Now we must go through some conceptual definitions. Especially we will have to define what a CDO is. Wang, Rachev and Fabozzi (2006) define a collaterized debt obligation as a security backed by a diversified pool of one or more kinds of debt obligations such as bonds, loans, credit default swaps (CDSs) or structured products (MSBs, ABSs) or other CDOs. It can be initiated only by a *sponsor*, which normally will be a bank, a non-bank financial institution or an asset management company.

The sponsor of the CDO must create a company, which is the SPV (Special Purpose Vehicle). This SPV is an independent entity, whose purpose is to isolate the CDO investors from the credit risk of the sponsor. The SPV then purchases debt obligations (bonds or loans) or sells CDSs to obtain credit risk exposure. The risk is transferred after by issuing debt obligations (which in this case would be tranches). Finally the investors in these tranches of the CDO have to final credit risk exposure to the underlying assets.

In the literature, for example Lucas, Goodman and Fabozzi (2006), these tranches are classified in three categories according to their subordinate levels; these are the subordinate/equity tranche, mezzanine tranches, and senior tranches. If the SPV of a CDO is the owner of the underlying obligations, this would be classified as a *cash* CDO. These, in turn can be of two types, which are the CBOs (Collateralized Bond Obligations) and CLOs (Collateralized Loan Obligations). The pool of the obligations of the former consists only of bonds, while the latter have only commercial loans in their pool.

Most of these definitions are not necessary for the purpose of this article and for the model to obtain but are chose for expositional convenience. To price the default correlation in tranches of a CDO, Hull, Predescu, and White (2005) propose the structural model. Wang, Rachev and Fabozzi (2006) also explore the model. This idea is based on the previously exposed Merton's [2] model and the extension made by Black and Cox (1976). As mentioned previously it is assumed that the value of a company follows a stochastic process, and if the value of the assets of the company go below a certain level (i.e. the barrier) the company will have to default.

## CDO pricing method (4.2)

In this model, we newly have to assume that there are N companies, and the value of a given company $i$ ($1 \leq i \leq N$) at time t is $V_t$. Also we assume that the risk-neutral diffusion process that value of the company follows is

$$(5) \quad dV_t = \mu_t V_t dt + \sigma_t V_t dz_t,$$

where $\mu_t$ is the expected growth rate of the value of the given company $i$, $\sigma_t$ is the volatility of the company $i$ and $z_t$ is a variable following a Weiner process. and also could be expressed like this

The (5) also could be expressed like this

$$(6) \quad dV_i = \mu_i V_i dt + \sigma_i V_i dX_i$$

We can label the barrier for company $i$, as $B_i$, so when the value of the assets of the company falls below this barrier, it defaults. Also we have to assume that $X_i(0) = 0$, without losing the generality.

After if we apply the Itô's Lemma, to the $\ln V_i$ we can show that

$$(7) \quad X_i(t) = \frac{\ln V_i(t) - \ln V_i(0) - (\mu_i - \sigma_i^2/2)}{\sigma_i}$$

Also with $B_i$, we have a barrier $B_i^*$ for $X_i$, as the following

$$(8) \quad B_i^* = \frac{\ln B_i - \ln V_i(0) - (\mu_i - \sigma_i^2/2)}{\sigma_i}$$

Here we have four unobservable parameters, $V_i(0)$, $H_i$, $\mu_i$ and $\sigma_i$, that can be expressed this way:

$$(9) \quad \beta_i = \frac{\ln H_i - \ln V_i(0)}{\sigma_i} \quad , \quad \gamma_i = -\frac{\mu_i - \sigma_i^2/2}{\sigma_i}$$

then we will have that

$$(10) \quad B_i^* = \beta_i + \gamma_i$$

So we can understand that when $X_i$ falls below the value of $B_i^*$ the company $i$ will have to default.

If we want to model the default correlation we will have to assume that each Weiner process $X_i$ follows a two component process. This consists of a common component $F_i$ and an idiosyncratic component $U_i$. This is expressed bellow as

$$(11) \quad dX_i(t) = \alpha_i(t)dF(t) + \sqrt{1 - \alpha_i(t)^2}\, dU_i(t),$$

where

the variable $\alpha_i$ can be a function of time or a stochastic, is used to control the two component process, and the Weiner processes $F_i$ and $U_i$ are uncorrelated with each other. In this model, the default correlation between the processes followed by the assets of the two companies $i_1$ and $i_2$ is $\alpha_{i1}\alpha_{i2}$.

This way Hull, Predescu, and White (2005) establish a structural model, with that advantages that it is a dynamic model and the correlation parameters can be estimated empirically.

# V. Conclusions

In previous sections we have exposed some works found in the literature on the structural model of default for a single company and the value of the debt issue. Also we have presented works on the structural model to price the default correlation in tranches of a CDO.

Using the tools and the assumptions from the Sections III and IV we can establish a framework that will allow us to create a different type of security. The function of which would be allowing all types of companies to free capital and to enter in temporal tax breaks from different types of taxes (VAT, corporate income taxes, etc).

It would be interesting to see the empirical results of this theoretical example bearing in mind that near 0, 0 and negative interest rates are prevailing in some parts of the world.

Finally, this type of contract has a double objective. First it could help companies to temporarily reduce taxes not related to their income taxes, and second, it could help them to free capital, i.e. taxes that are related to income. The second objective would especially apply for those companies which have to disburse the tax amount before receiving the income amount.